\documentclass[10pt,conference]{IEEEtran}

\usepackage{graphicx}
\usepackage{epsf}
\usepackage{verbatim}
\usepackage{algorithmic}
\usepackage[ruled,vlined]{algorithm2e}
\usepackage{url}
\usepackage{makecell}
\usepackage{alltt}
\usepackage{listings}
\usepackage{longtable,lscape}
\usepackage{slashbox,multirow}
\usepackage{colortbl}
\usepackage{mathrsfs}
\usepackage[labelfont=bf]{caption}
\usepackage{subcaption}
\usepackage{enumerate}
\usepackage[all]{xy}
\usepackage{float}
\usepackage{threeparttable}
\usepackage{mathrsfs}
\usepackage{balance}

\usepackage{amssymb}
\usepackage{booktabs}
\usepackage{xcolor}
\usepackage{xspace}
\usepackage{hyperref}

\definecolor{codeviolet}{rgb}{0.65,0.11,0.36}
\newcommand{\tool}{\textsc{AutoUpdate}\xspace}

\newcommand{\CodeIn}[1]{{\footnotesize\texttt{#1}}}

\newcommand{\Comment}[1]{}

\newenvironment{SmallOut}{\begin{small}}{\end{small}}

\IEEEoverridecommandlockouts
\usepackage{cite}
\usepackage{amsmath,amssymb,amsfonts}
\usepackage{algorithmic}
\usepackage{graphicx}
\usepackage{textcomp}
\usepackage{xcolor}
\def\BibTeX{{\rm B\kern-.05em{\sc i\kern-.025em b}\kern-.08em
    T\kern-.1667em\lower.7ex\hbox{E}\kern-.125emX}}
\begin{document}

\title{Migrating Client Code without Change Examples}

\author{
\IEEEauthorblockN{Hao Zhong}
\IEEEauthorblockA{\textit{Department of Computer Science and Engineering}}
\textit{ Shanghai Jiao Tong University, China}\\
zhonghao@sjtu.edu.cn
\and
\IEEEauthorblockN{Na Meng}
\IEEEauthorblockA{\textit{Department of Computer Science}}
\textit{Virginia Polytechnic Institute and State University, USA}\\
nm8247@cs.vt.edu
}

\maketitle

\begin{abstract}
API developers evolve software libraries to fix bugs, add new features, or refactor code. To benefit from such library evolution, the programmers of client projects have to repetitively upgrade their library usages and adapt their codebases to any library API breaking changes (\emph{e.g.}, API renaming). Such adaptive changes can be tedious and error-prone.
Existing tools provide limited support to help programmers migrate client projects from old library versions to new ones. For instance, some tools extract API mappings between library versions and only suggest simple adaptive changes (\emph{i.e.}, statement updates); other tools suggest or automate more complicated edits (\emph{e.g.}, statement insertions) based on user-provided exemplar code migrations. 
However, when new library versions are available, it is usually cumbersome and time-consuming for users to provide sufficient human-crafted samples in order to guide automatic migration.

In this paper, we propose a novel approach---\tool---to further improve the state of the art.
Instead of learning from change examples, we designed \tool to automate migration in a compiler-directed way. Namely, given a compilation error triggered by upgrading libraries, \tool exploits 13 migration operators to generate candidate edits, and tentatively applies each edit until the error is resolved or all edits are explored.
 We conducted two evaluations. The first evaluation involves migrating 371 tutorial examples between versions of 5 popular libraries. \tool reduced migration-related compilation errors for 92.7\% of tasks. It eliminated such errors for 32.4\% of tasks,  and 33.9\% of the tasks have identical edits to manual migrations. In the second evaluation, we applied \tool to migrate two real client projects of \CodeIn{lucene}. \tool
 successfully migrated both projects, and the migrated code passed all tests.
\end{abstract}


\section{Introduction}
\label{sec:intro}
Application programming interface (API) libraries provide reusable APIs and are widely used in software development. As API developers constantly repair bugs and implement new features, it is likely that client code can benefit from new versions of libraries. Despite of the potential benefits, it is usually tedious and error-prone for client-code programmers to upgrade their projects, because new libraries contain many \textbf{API breaking changes}~\cite{sawant2016reaction,kula2018developers}. For example, when an old API \CodeIn{foo()} does not exist in the new library version, programmers have to adapt client code after library upgrade, and replace any usage of \CodeIn{foo()} with that of new API(s).


\textbf{The challenges.} Migrating a project to call a newer library has the following challenges:

\emph{1. API mappings alone are insufficient to migrate clients.} Lamothe and Shang~\cite{lamothe2018EUA} observed more complicated modifications than renaming code names and changing parameters during migration. Some such modifications were observed by other researchers. For example, Dig and Johnson~\cite{dig2006apis} identified complicated API changes such as new method contracts and new interfaces. It is difficult to define them as mappings.

\emph{2. There are multiple ways to migrate an API call.} As one API element (a type, a method, or a field) can be split to multiple ones, the call sites of an API can have different ways to be migrated.

\emph{3. Many replacements have no documentation.} When APIs become deprecated, their replacements can be not documented. The prior studies report that 26\%~\cite{lamothe2018EUA} to 60\%~\cite{brito2016developers} replacements are not described in documents.


\emph{4. Resolving a compilation error can lead to more errors.} For example, after we replace a missing type with a new type, more incompatible method errors are introduced, because the new methods change their parameters.

{\textbf{The state of the art.} To reduce the manual effort of client code migration,
researchers proposed various approaches (Section~\ref{sec:related}). Beased on their migration capabilities, these approaches fall into either of the following categories:

\emph{1. Mining API mappings.} API mappings provide hints to replace old API calls with their replacements. Nguyen \emph{et al.}~\cite{nguyen2015graph,nguyen2014statistical,nguyen2013lexical} combine graphs with statistical learning to mine API mappings from two versions of clients. Chen \emph{et al.}~\cite{chen2019mining} combine the source code and comments in clients to mine API mappings. When libraries are updated, some APIs stay unchanged. Wu \emph{et al.}~\cite{wu2010aura} call such unchanged APIs as anchors, and analyze how the anchors are called by other changed APIs to obtain more mappings. Meng \emph{et al.}~\cite{meng2012history} compare the revisions of libraries to mine API mappings. Lamothe and Shang~\cite{lamothe2018EUA} criticize that API mappings alone are insufficient to migrate clients. As these approaches focus on only one aspect of migrating clients, researchers either do not migrate clients (\emph{e.g.}, \cite{wu2010aura,meng2012history}), or they admit that mined mappings only reduce migration efforts (\emph{e.g.}, \cite{zhong2010mining}) in their evaluations. Here, two versions of libraries and clients can be considered as code change examples.


\emph{2. Migrating with change examples.} Based on the types of change examples, these approaches can be roughly divided into two research lines. The first line of approaches analyzes the examples of library changes, and infer the corresponding adaptive changes in client code. For example, Xing and Stroulia~\cite{xing2007api} break the UMLs of two libraries down, and infer their corresponding code changes on clients. The second line of approaches analyzes the examples of client code changes, and infer transformations that can be applied to new code locations. For example, Meng \emph{et al.}~\cite{Meng:LASE} learn an edit script from a migration example, and apply the script to similar code locations. As migration examples are expensive to be manually constructed, Fazzini \emph{et al.}~\cite{fazzni2019issta} propose an approach to search for such examples from a code repository.

In summary, the state-of-the-art approaches mainly use change examples to resolve the above challenges, but it is very expensive to construct such examples.

\textbf{Our contributions} are as follows:

\emph{1. A novel research direction of building complicated migration edits from simple ones}. Instead of breaking complicated changes, we construct complicated changes from simple edits. To achieve this goal, we reduce the construction into an optimization problem of searching for a best combination of simple edits. The target of an optimization problem~\cite{beck1964linear} is to search for the best solution from all feasible solutions.

\emph{2. The first approach, called \tool, that attacks the migration problem with synthesis}. To determine which one is the best, there must be an algorithmic way of measuring the fitness values of solutions. Our fitness function calculates the number of compilation errors, and our optimized solution is a migrated project whose compilation errors are all resolved.

Our evaluation results show that \tool made the following highlights:

\emph{1. Two real projects were successfully migrated, and the migrated projects passed all their test cases.}
\tool migrated two open-source Java projects---\CodeIn{ESA}~\cite{esa} and \CodeIn{FleaDB}~\cite{fleadb}---from an old version of the \CodeIn{lucene} library to a new version, although the migration tasks involved complex modifications.

\emph{2. The compilation errors of 92.7\% tasks were reduced, and those of 32.4\% tasks were eliminated.} For each version of another five libraries, we created a migration task, in which the API tutorial examples of \CodeIn{v$_{old}$} were added to the source scope and the binary files of \CodeIn{v$_{new}$} (\emph{i.e.}, jar files) were added to the library scope. Each task is challenging because the examples call many APIs. In total, we used \tool to migrate 371 tasks. \tool reduced the compilation errors in 92.7\% of tasks, and removed the compilation errors in 32.4\% of tasks.

\emph{3. 39.3\% migrated tasks are identical to manual updates.} We compared the modifications of \tool with manual updates. Our results show that (1) in 39.3\% migration tasks, \tool produces migrations that are identical to manual ones, and (2) the compilation errors in 54.1\% migration tasks are never resolved by programmers, but were fully migrated by \tool.



\section{Motivating Example}
\label{sec:example}
In this section, we construct a migration task to show the challenges of code migration, and to illustrate the benefits of \tool. More compilation errors are introduced, when a target library becomes more different from the original one. To construct a more challenging task, we compile an API tutorial example (\CodeIn{simple authentication}) of cassandra \CodeIn{1.0.0} with the binary files of cassandra \CodeIn{3.0.0}. In total, the task has 20 compilation errors. We analyzed these errors, and we found the instances of all our listed four challenges:

\emph{Challenge 1. More edits than replacements.} We found that the example implements the \CodeIn{IAuthenticator} interface. Compared with the one of \CodeIn{1.0.0}, the interface of \CodeIn{3.0.0} defines five additional methods. As the example project does not implement the five methods, the compiler produces error messages, and one such message is ``The type SimpleAuthenticator must implement the inherited abstract method I\-Au\-then\-ti\-ca\-tor.\-setup()''. It is insufficient to handle this problem with only replacements. To handle this challenge, \tool generates stub methods for each needed method.

\emph{Challenge 2. More than one-to-one mappings.} The example calls the \CodeIn{FBUtilities.\-hex\-To\-Bytes} method.
In \CodeIn{3.0.0}, the \CodeIn{FBUtilities} class is split into a new \CodeIn{FBUtilities} class and a \CodeIn{ByteBufferUtil} class, and the \CodeIn{hexTo\-Bytes} method is moved to the later one. As a result, the type is mapped to two new types. To handle this type of problems, \tool implements more migration operators than renaming API calls and enumerates other operators until one works. In particular, \tool finds the replacement method, because the method has the most similar name and the identical parameter list to the original one.

\emph{Challenge 3. Undocumented replacements.} As introduced in the previous paragraph, the \CodeIn{hex\-To\-Bytes} method is moved to a new class. However, we find that neither the documentation of \CodeIn{1.0.0} nor the documentation of \CodeIn{3.0.0} mention that the method is moved. As cassandra provides thousands of API methods, it is difficult for programmers to manually locate the replacement of this method. To handle this challenge, we implement a tool to automatically mine the mappings from API documentation, and even without mined mappings, \tool implements other operators to update code. In this example, \tool finds the correct replacement by matching their method names and parameter types.

\emph{Challenge 4. More compilation errors after one is resolved.} This example project implements the \CodeIn{IAuthority} interface. In \CodeIn{3.0.0}, the interface is merged with \CodeIn{IAuthenticator}, and the new interface is named as \CodeIn{IAuthenticator}. As JDT does not find the \CodeIn{IAuthority} interface, it reports a compilation error. This problem cannot be resolved by a single renaming action. After we replace the interface with the correct one, JDT reports more compilation errors. The new interface defines more methods than the old one. After the old class is replaced, the compiler reports that the added methods must be implemented. To fix the problem, it must combine multiple migration operators with proper steps. \tool uses a migration algorithm to guide its migration process.

When programmers update the example project, they also have to resolve all the compilation errors. As resolving one problem can lead to more errors, we have to modify more locations than the initially reported 20 compilation errors. As migrating this example needs more than mappings, only mined mappings are insufficient to migrate the example. Meanwhile, it is difficult to find its change examples, and writing change examples is tedious. \tool spent only several minutes to migrate the example, and its solution is identical to our manual migration. The results show two benefits:

\emph{Benefit 1. \tool updates client code in a fully automatic manner.} The prior approaches~\cite{henkel2005catchup,xing2007api,meng2012history,wu2010aura} mine API mappings, but Lamothe and Shang~\cite{lamothe2018EUA} show that it is insufficient to migrate clients with only replacements. \tool handles more issues than replacing APIs with its migration operators (Section~\ref{sec:approach:action}).

\emph{Benefit 2. \tool is able to migrate code locations that require multiple and complicated modifications without change examples.} Lamothe and Shang~\cite{lamothe2018EUA} complain that some changes in migration are complicated (\emph{e.g.}, many-to-many mappings), and suspect that such migrations must be learnt from examples. Although \tool does not require migration samples, it overcomes the challenges, since it generates complicated changes.

In summary, \tool presents a way to resolve all the four challenges in Section~\ref{sec:intro} without any change examples.

\section{Research Roadmap}
\label{sec:road}
In this section, we analyze our research roadmap.


\textbf{The comparison to prior approaches.} For the simplicity of our discussion, we use a function, $f(x)\rightarrow y$, to denote an approach, where $x$ is the input, $y$ is the output, and $f$ is its technique. After this approach is proposed, an researcher can propose their $f^\prime$ to attack the identical problem. To show that $f^\prime$ is better than $f$, the researcher shall present empirical evidences. In this scenario, the researcher shall conduct controlled experiments to compare $f$ with $f^\prime$. In this controlled experiment, the researcher shall align the input $x$; switch the treatments between $f$ and $f^\prime$; and compare their output $y$. Although we agree that controlled experiments are the foundation of the scientific research~\cite{baker1984controlled}, it is infeasible or unnecessary to conduct such an experiment in some cases. For example, if a novel approach takes quite different inputs from the prior ones, it becomes infeasible to align the inputs. As a result, even if $f$ produces better outputs than $f^\prime$ does, it cannot conclude that $f$ is better than $f^\prime$. In this case, it is infeasible and unnecessary to conduct an controlled experiment to show the improvements. For example, most approaches mine specifications from clients~\cite{robillard2012automated}, but many APIs do not have sufficient clients for mining~\cite{zhong2017empirical}. Zhong \emph{et al.}~\cite{zhong09:doc2spec} proposed the first approach to mine specifications from API documents. As it is infeasible to align their inputs to the prior approaches, they did not conduct any controlled experiments in their evaluations. As we remove change examples from the inputs of migrating clients, it becomes infeasible to align the inputs of \tool with the prior approaches. Even if \tool completes more migration tasks than a prior approach, a critical researcher will complain that change examples are insufficient, but including change examples is already unfair to \tool, since it does not need any change examples at all. Indeed, eliminating the effort to preparing change examples is our major contribution, but the effort is not even considered if a compared approach uses such examples. As a result, in our evaluation, we did not compare with any prior approaches.

\textbf{The comparison of follow-up approaches.} As there is no constraints on how a library can evolve, the deltas from a library to its future versions are not a constant set. As a result, it is infeasible to derive a complete set of operators, and the quests for better operators and migration algorithms are endless. From the research angle of operators, we envisage that researchers can obtain more effective migration operators from four sources: (1) After the prior approaches (\emph{e.g.}, \cite{jiang2019inferring,Meng:LASE}) learn transformation patterns from change examples, it becomes feasible to mine those common ones as migration operators; (2) it is feasible to infer better migration operators from the prior empirical studies on API breaking changes and behavioral backward incompatibilities~\cite{raemaekers2017semantic,jezek2017api,mostafa2017experience}; (3) As Cossette and Walker~\cite{cossette2012seeking} show that on average only 20\% code changes can be realized by the prior migration approaches, it can be feasible to design migration operators based on their observed failure cases; (4) software engineering thesauruses~\cite{beyer2015synonym,wang2012inferring} are useful to design better migration operators; and (5) the code change types identified by prior studies~\cite{fluri2008discovering} are useful to design migration operators. Still, even if operators are sufficient, their migration capability can be unleashed with more advanced algorithms (\emph{e.g.}, the swarm algorithm~\cite{schutte2004parallel}) can be more effective to guide the migration process than ours. Our dataset can be released to construct a benchmark for follow-up researchers, and on this benchmark, they can conduct controlled experiments to show their improvements. Furthermore, they shall present their migration capability beyond only benchmarks. 

\textbf{The good intentions of modifications.} As an automatic approach, we need a measure to guide the migration process, and it is natural to compilation errors as the measure. Indeed, when programmers manually update code, as the first step, they also have to manually resolve compilation errors. Researchers~\cite{qi2014strength,qi2015issta} show that it is feasible to cheat an automatic measure. A recent study~\cite{wu2021bug} shows that even experienced programmers may not identify vulnerabilities in pull requests, if they are maliciously and deliberately embedded. To ensure the good intentions of modifications, the results of an automatic measure (\emph{e.g.}, fewer compilation errors) alone are insufficient to prove an improvement over ours. We suggest that researchers shall describe their operators clearly, and release their changed code and logs as we did, so that other researchers can manually inspect and confirm the reliability of their improvements.

\textbf{The work after migration.} Even if all modifications are in good intentions, migrated code can still be buggy. 
Zhong \emph{et al.}~\cite{zhong2013exposing} show that mapped APIs have semantics differences. For example, two mapped APIs can produce different output when inputs are corner cases (\emph{e.g.}, \CodeIn{null}). Although such differences can cause bugs in migrated code, to the best of our knowledge, no migration tools guarantee the semantic correctness of migrated code. Still, only after the compilation errors of a migrated project are fully removed, researchers can define other fitness functions to further improve the migrated project. For example, automatic program repair and repairing corrupt data (\emph{e.g.}, \cite{weimer2009automatically}) use the number of passing test cases to guide their repair process. If a project still has compilation errors, the above approaches cannot remove its bugs, since test cases will fail on code with compilation errors.

\section{Approach}
\label{sec:approach}
To reduce compilation errors in migrated code, \tool prepares the inputs for updating (Section~\ref{sec:approach:input}), and defines a set of migration operators to update code (Section~\ref{sec:approach:action}). The process is guided by our migration algorithm (Section~\ref{sec:approach:guide}).

\begin{table*}[t]
\centering
\caption{Our migration operators.}\vspace*{-2ex}
\begin{SmallOut}
\begin {tabular} {|c|l|l|l|r|r|r|r|}
\hline
 \multicolumn{1}{|c|}{Category} &   \multicolumn{1}{|c|}{Id}  & \multicolumn{1}{|c|}{Migration action} &  \multicolumn{1}{|c|}{Target compilation error} \\
\hline\hline
   \multirowcell{4}{Missing API \\elements} &    MA1   & Replacing undefined API elements with mappings & undefined types, methods, and variables \\
\cline{2-4}
    &MA2   & Replacing undefined API elements with compatible ones & undefined methods and variables \\
\cline{2-4}
    &MA3   & Replacing undefined fields with getters/setters & undefined fields \\
\cline{2-4}
    &MA4   & Replacing undefined constructors with creators & undefined constructors \\
\hline\hline
    \multirowcell{4}{Incompatible \\API elements} &   MA5   & Generating explicit conversions & class hierarchy changes (incompatible types)\\
\cline{2-4}
  &  MA6   & Reducing or swapping method parameters & incompatible methods \\
\cline{2-4}
   & MA7   & Replacing static calls with instance calls & undefined and incompatible methods \\
\cline{2-4}
    &  MA8   & Exploring declared fields and methods & incompatible actual parameters\\
\hline\hline
    \multirowcell{3}{More or fewer\\ API calls}  &  MA9   & Generating method stubs & unimplemented methods \\
\cline{2-4}
   & MA10   & Handling exceptions & unhandled exceptions \\
\cline{2-4}
   & MA11  & Removing API calls & undefined methods and variables \\
\hline\hline
  \multirow{2}*[-2pt] {Other issues}  & MA12  & Resolving ambiguous types & ambiguous types \\
\cline{2-4}
   & MA13  & Replacing invisible fields with getters and setters & invisible fields \\
\hline
\end{tabular}\\
\end{SmallOut}\vspace*{-1ex}
\label{table:ra}
\end{table*}

\subsection{Preparing Inputs for Updating}
\label{sec:approach:input}
As the first step, \tool prepares the inputs such as error locations, API documentation, and API mappings.

\emph{1. Extracting error locations with JDT.} An error location is a code element, which introduces a compilation error after API libraries are updated. \tool uses JDT~\cite{jdt} to extract compilation errors, since we focus on updating Java code and JDT is the built-in compiler of the Eclipse IDE. When JDT locates a compilation error, it reports the code range of the error. Based on each code range and its corresponding Abstract Syntax Tree (AST), \tool locates its code element with errors.

Besides code ranges, JDT reports the types of compilation errors. For example, if it fails to resolve a type named \CodeIn{A}, JDT reports its error message: ``\CodeIn{A} cannot be resolved to a type.'' \tool parses these compilation errors to determine their error types. For each type of compilation errors, \tool applies corresponding migration operators as defined in Section~\ref{sec:approach:action}.

\emph{2. Extracting API elements and their mappings from their API documentation.} As Lamothe and Shang~\cite{lamothe2018EUA} report that API documents describe most API replacements, we choose to extract API elements and their mappings from their documents. In Java, as API documents are generated by the Javadoc tool~\cite{javadoc}, they follow strict formats, and \tool extracts API elements (\emph{i.e.}, types, methods, and fields) by matching these formats. After the elements are extracted, \tool uses the Hungarian algorithm~\cite{kuhn1955hungarian} to extract their mappings. The Hungarian algorithm is a classical algorithm to extract the best mappings between two sides of items. The algorithm needs a function to define the distance between two items, and it finds mappings that can minimize the overall distance. For two API elements of the same type, \tool uses the Levenshtein edit
distance of their full code names to calculate their distance. One of our migration operators (\emph{i.e.}, MA1 in Section~\ref{sec:approach:action}) uses mappings to update client code. Although the algorithm mines only one-to-one mappings, \tool can migrate code that requires many-to-many mappings, since it can combine simple edits to complicated ones (see Section~\ref{sec:approach:guide}). Compared with other API mapping approaches(\emph{e.g.}, \cite{chen2019mining}), our extraction is simple. \tool can achieve even better results, if follow-up researchers use more advanced techniques to mine API mappings.

\subsection{Resolving Compilation Errors}
\label{sec:approach:action}
Table~\ref{table:ra} shows the mappings between our migration operators and their target compilation errors. We implement our migration operators on Spoon~\cite{pawlak:hal-01169705}, which is a library that allows the analysis and modifications of Java code. Although updating libraries introduces compilation errors in source files, it does not introduce syntactical errors. As a result, it is still able to generate correct modifications.

We design these migration operators with the following steps. First, we tried to update libraries of several projects, and collected their compilation errors. Second, we tried to recreate the challenges that are mentioned in the prior studies~\cite{vstrobl2013migration,dagenais2011recommending,dig2006automated,henkel2005catchup,nguyen2016api,lamothe2018EUA}, and collected their compilation errors. After we collected as many as possible compilation errors, we analyzed how to resolve each type of compilation errors, and designed their migration operators. In particular, by tentatively replacing API(s) based on each type of mappings, we check whether the related compilation issue(s) are resolved, and thus construct our migration operators.

We can obtain a complete set of migration operators, if we break changes into the finest edits. After all, all modifications are the additions and deletions on a limited number of letters, but such a complete set of operators makes the search space extremely large. A practical tool must balance between the effectiveness and performance. As the first approach in this research line, we cannot guarantee that our migration operators are complete. Still, our positive results show a novel and effective way to resolve migration issues (see Section~\ref{sec:evaluation}), and we further discuss this issue in Section~\ref{sec:discuss}.

\subsubsection{Missing API elements} This category includes undefined types, undefined methods, undefined fields, invisible code elements, and incompatible methods. \tool implements the following migration operators to resolve this type of compilation errors:

\emph{MA1. Replacing undefined API elements with mappings.} As introduced in Section~\ref{sec:approach:input}, we have mined mappings from API documents. For an undefined element, MA1 queries mined API mappings using the full name of the code element as the keyword. If a replacement is found, MA1 replaces the undefined element with the replacement. For the example in Section~\ref{sec:example}, \tool replaces the missing \CodeIn{IAuthority} interface with the \CodeIn{IAuthenticator} interface, since the mapping \CodeIn{IAuthority}$\rightarrow$\CodeIn{IAu\-thenticator} is extracted.


\emph{MA2. Replacing undefined API elements with compatible ones.} If a method or a field is undefined, MA2 searches the code elements of the newer library, to locate its compatible matches. A compatible code element has a similar name, and introduces no more compilation errors, after it replaces the missing one. Here, we consider only the top ten similar items, and the similarity is the inverse of the Levenshtein edit distance between code names. For example, the \CodeIn{FBUtilities} type in \CodeIn{1.0.0}  is split into the \CodeIn{FBUtilities} type and the \CodeIn{ByteBufferUtil} type in \CodeIn{3.0.0}. As \tool mines only one-to-one mappings, the old \CodeIn{FBUtilities} class is mapped to the new \CodeIn{FB\-Utilities} class. The mined mapping is insufficient to resolve the compilation error. After searching the methods in \CodeIn{3.0.0}, MA2 locates the \CodeIn{ByteBufferUtil.hexTo\-Bytes} as the correct replacement.


\emph{MA3. Replacing undefined fields with getters or setters, and vice versa.} If a field is missing, MA3 will replace it with its getters or setters. When this happens, a field name is typically similar to the names of its getter and setter. For example, the sample of Section~\ref{sec:example} uses the \CodeIn{username} field of the \CodeIn{Au\-then\-ti\-catedUser} type. In \CodeIn{3.0.0}, this public field is changed to private. To handle the problem, \tool replaces the field with the \CodeIn{getName} method. Here, it first removes \CodeIn{get} or \CodeIn{set} from getters and setters. After that, it calculates the Levenshtein edit distances between the remaining method names and the field names, and select the one with the least distance. Meanwhile, if a getter or a setter of a field is deleted, \tool will try to replace it with the field. 

\emph{MA4. Replacing undefined constructors with creators.} To implement a Factory design pattern, API developers can delete or hide the constructors of a class, and implement creators for the class. MA4 will replace such deleted and hidden constructors with their creators. For example, the samples of \CodeIn{cassandra} \CodeIn{0.8.8} call the constructor of the \CodeIn{ColumnFamily} type, but the constructor is hidden in later versions. Instead, \CodeIn{ColumnFamily} implements a set of static methods to create the type. \tool replaces the constructor with these creators to resolve the compilation error. 

\subsubsection{Incompatible API elements} This category includes conversion errors, and incompatible/undefined methods. \tool implements the following migration operators:

\emph{MA5. Generating explicit conversions.} For type conversion errors, MA5 adds a cast expression to explicitly convert it. For example, the migration task in Section~\ref{sec:example} involves an assignment:

\begin{lstlisting}
authorized = Permission.ALL;
\end{lstlisting}

JDT reports a type mismatch error. To resolve the problem, \tool adds an explicit cast:

\begin{lstlisting}
authorized = (EnumSet<Permission>) Permission.ALL;
\end{lstlisting}


\emph{MA6. Reducing or swapping method parameters.} When the signature of a method is changed, JDT can report incorrect overridden methods if the method is overridden, or incompatible methods if the method is directly called. Lamothe and Shang~\cite{lamothe2018EUA} also notice the problem, and introduce a case  (\emph{i.e}, the \CodeIn{queue(Bytebuffer, int)} method), in which the second actual parameter must be removed. Following their suggestions, \tool compares the new signature of its actual parameters of a call site. If a parameter is deleted, \tool removes its corresponding actual parameter from the call site. Alternatively, if the parameter order is changed, \tool reorders the parameters based on parameter types. 

\emph{MA7. Replacing static calls with instance calls, and vice versa.} If a static method is deleted, MA7 will replace it with instance methods, and vice versa. For example, the samples of \CodeIn{cassandra} \CodeIn{0.8.0} call \CodeIn{Byte\-Buf\-fer\-Util\-.bytes(key)} to obtain the bytes of \CodeIn{key}, but later versions delete the static method. The type of \CodeIn{key} is \CodeIn{Text}, and in the later versions, an instance method \CodeIn{getBytes()} is added to \CodeIn{Text} to obtain the bytes of \CodeIn{key}. To resolve the error, \tool replaces the static method call with \CodeIn{key.getBytes()}. 

\emph{MA8. Exploring declared methods and fields of an incompatible type.} If a method has an actual parameter whose type is \CodeIn{Type1} but the desirable one is \CodeIn{Type2}, MA8 explores all the fields and methods that are declared by \CodeIn{Type1}. If the type of a field is \CodeIn{Type2}, MA8 modifies the actual parameter to reference this field. Similarly, if the return type of a method is \CodeIn{Type2}, MA8 modifies the actual parameter to call this method.

\subsubsection{More or fewer API calls} Programmers can add or delete API calls to resolve compilation errors during migration. \tool imitates the process to resolve these errors.

\begin{algorithm}[t]
\footnotesize
\begin{algorithmic}[1]
\REQUIRE~~\\
$p$ is the project whose library is updated. \\
\ENSURE~~\\ $p^\prime$ is the migrated project.
 \STATE errors$\leftarrow$ compile(p); pool$\stackrel{add}{\longleftarrow}$ p; nobetter$\leftarrow$ 0; generation$\leftarrow$ 0;
 \WHILE {errors $\neq \emptyset$  and nobetter$<$10 and generation$<$100}
    \STATE nextpool$\leftarrow \emptyset$ ;
    \FOR{p $\in$ pool}
      \STATE errors$\leftarrow$ compile(p);
      \FOR{error $\in$ errors}
          \STATE op$\gets$ operator(error); solution$\gets$ apply(op, p);
          \STATE nextpool$\stackrel{add}{\longleftarrow}$ solution;
      \ENDFOR
    \ENDFOR
      \STATE best$\gets$ best(pool); nextbest$\gets$ best(nextpool);
      \IF {compile(nextbest)$<$compile(best)}
          \STATE nobetter$\gets$nobetter+1;
         \ELSE
           \STATE errors$\gets$compile(nextbest);\STATE nobetter$\gets$0;$p^\prime\gets$ nextbest;
      \ENDIF
      \STATE generation$\gets$ generation+1; pool$\stackrel{top 10}{\longleftarrow}$ nextpool;
 \ENDWHILE
\end{algorithmic}
\caption{the \CodeIn{migration} Algorithm}%
\label{alg:mig}
\end{algorithm}

\emph{MA9. Generating method stubs.} Suppose that a client-code class (\CodeIn{CC}) extends an API abstract class (\CodeIn{AC}) and implements an interface (\CodeIn{AI}), and in a follow-up version, a new abstract method is added to \CodeIn{AC}, or a new method is added to \CodeIn{AI}. In both cases, \CodeIn{CC} must implement the newly added method to avoid compilation errors. If JDT reports that a type (\CodeIn{C}) does not implemented a method (\CodeIn{m}), MA8 searches the super classes of \CodeIn{c} for the signature of \CodeIn{m}. With the found signature, \tool generates a method stub for \CodeIn{C}. For example, the example of \CodeIn{0.8.0} has a \CodeIn{SimpleAuthenticator} class, which implements the \CodeIn{IAuthen\-ti\-ca\-tor} interface. A later version adds more methods to the interface, but the class does not implement the added methods. As a result, the migrated example produces a compilation error: ``The type \CodeIn{SimpleAuthenticator} must implement the inherited abstract method \CodeIn{IAuthen\-ti\-ca\-tor.\-re\-quire\-Au\-then\-ti\-ca\-tion()}''. To resolve the problem, \tool generates a method stub inside \CodeIn{Simple\-Authen\-ti\-cator}:


\begin{lstlisting}
public boolean requireAuthentication() {
//todo: Please implement the method.
  return false;}
\end{lstlisting}

After the method stubs are generated, programmers have to implement its method body. As this migration operator indicates, programmers have to implement new code, when they migrate code to call newer APIs. We further discuss this issue in Section~\ref{sec:discuss}.

\emph{MA10. Handling exceptions.} If the thrown exceptions of a method are modified, JDT can report exception-related errors.
To remove such errors, MA9 adds \CodeIn{throw} expressions or \CodeIn{try-catch} statements at error locations. For example, the \CodeIn{keyspace} method of \CodeIn{cassandra} \CodeIn{1.2.0} throws two exceptions, but later versions throw three exceptions. As a result, when we migrate the examples of \CodeIn{cassandra} \CodeIn{1.2.0} to later versions, JDT reports a compilation error that an exception is not handled. \tool modifies the corresponding client code to handle the new exception.

\emph{MA11. Removing API calls.} Programmers can remove API calls, since their corresponding API elements are removed. For example, the examples of \CodeIn{poi} \CodeIn{3.16} call the following method:

\begin{lstlisting}
boldFont.setBoldweight(HSSFFont.COLOR_NORMAL.BOLDWEIGHT_BOLD);
\end{lstlisting}

As the later versions delete \CodeIn{BOLDWEIGHT\_BOLD}, the above code produces a compilation error during migration. We found that programmers simply delete the above code line to resolve the compilation error. This migration operator imitates this process. To reduce the possibility of removing useful code lines, \tool removes at most one line each time. As removing useful code lines introduces compilation errors, the guidance algorithm of \tool will penalize the abuse of this migration operator.

\begin{table}[t]
\centering
\caption{The subjects.}\vspace*{-2ex}
\begin{SmallOut}
\begin {tabular} {|c|r|r|r|r|r|r|r|}
\hline
    Name  & \multicolumn{1}{|c|}{FV} &  \multicolumn{1}{|c|}{LV} & \multicolumn{1}{|c|}{TV} & \multicolumn{1}{|c|}{LOC}\\
\hline\hline
      \CodeIn{accumulo} & 1.3.6 & 1.9.2 & 29    & 81,949  \\
\hline
    \CodeIn{cassandra} & 0.8.0 & 3.11.2 & 162   & 33,471  \\
\hline
    \CodeIn{karaf} & 1.6.0 & 4.2.3 & 72    & 16,618  \\
\hline
    \CodeIn{lucene} & 1.9.0 & 7.4.0 & 92    & 63,636  \\
\hline
    \CodeIn{poi}   & 3.0   & 4.0.1 & 21    & 102,901  \\
\hline
\end{tabular}\\
\end{SmallOut}\vspace*{-1ex}
\label{table:subject}\vspace*{-1ex}
\end{table}
\subsubsection{Other issues} \tool implements several migration operators to handle other issues of updating client code.

\emph{MA12. Resolving ambiguous types.} This migration operator is designed to resolve ambiguous types. If JDT reports that a type is ambiguous, MA12 generates a precise \CodeIn{import} statement to resolve the problem. As \tool works in a try-and-validate manner, it does not have to determine which type is the correct type. Instead, it generates a solution for each ambiguous type. For example, the examples of \CodeIn{cassandra} \CodeIn{1.2.0} import all the types of two packages:

\begin{lstlisting}
import org.apache.cassandra.db.*;
import org.apache.cassandra.thrift.*;
\end{lstlisting}

As the later versions implement two \CodeIn{ConsistencyLevel} types under the above packages, JDT reports that the type is ambiguous. \tool replace the above statements with:

\begin{lstlisting}
import org.apache.cassandra.db.ConsistencyLevel;
\end{lstlisting}

\tool generates a solution for each possible type. In the other solution, it replaces the above statement with the \CodeIn{Consistency\-Level} type under the \CodeIn{thrift} package. This solution is discarded, since it introduces more errors.

\emph{MA13. Replacing invisible fields with getters and setters.} An updated library can hide a visible field. In such cases, JDT reports that fields are invisible. To handle these errors, MA12 tries to replace such field accesses with getters and setters. For example, the examples of \CodeIn{poi} \CodeIn{3.15} have the following code:

\begin{lstlisting}
 if(attachment.attachLongFileName != null) {
\end{lstlisting}

The later versions hide the \CodeIn{attachLongFileName} field, and implement a getter called \CodeIn{getAttachLongFileName()}. \tool replaces the above field access with the corresponding getter to resolve this issue. This operator is similar to MA8. We implement a separate operator, since their error types are different. 


\subsection{Guiding the Migration Process}
\label{sec:approach:guide}
As introduced in the beginning of this section, the purpose of \tool is to reduce the compilation errors that are introduced in code migration. We reduce this purpose to a search problem, in which we minimize the compilation errors of a client whose libraries are updated. Genetic algorithms~\cite{whitley1994genetic} are widely used to solve optimization problems. Algorithm~\ref{alg:mig} shows our algorithm to guide the migration process, and it is inspired by genetic algorithms.

In our algorithm, Line 1 initiates our variables, and Line 2 shows our termination condition. It stops when (1) a solution without compilation errors is synthesized; (2) no better solutions are found in the recent ten generations; or (3) it fails to resolve all the compilation errors within one hundred generations. For each solution in our pool, Line 5 compiles it to collect its compilation errors. For each compilation error, Line 7 selects a migration operator, and applies the operator to generate a new solution. If multiple migration operators are feasible, \tool randomly picks one of them. Line 8 adds new solutions to a new pool. Line 11 obtains the best solutions of both pools. Line 12 compares the best solutions from the two pools. If the new pool does not contain a better solution, Line 13 increases \CodeIn{nobetter}. If it contains, Line 16 resets the variable and assigns the better solution to $p^\prime$. Line 18 increases \CodeIn{generation}, and selects the top ten unique solutions from the new pool. It determines that two solutions are duplicated, if their compilation errors are identical.



\begin{table}[t]
\centering
\caption{The examples of cassandra \CodeIn{0.8.0-beta1}.}\vspace*{-1ex}
\begin{SmallOut}
\begin {tabular} {|c|r|r|r|r|r|r|r|}
\hline
    Name  & \multicolumn{1}{|c|}{File} &  \multicolumn{1}{|c|}{LOC} & \multicolumn{1}{|c|}{Class} & \multicolumn{1}{|c|}{Method} & \multicolumn{1}{|c|}{Field}\\
\hline\hline
    client & 1     & 196   & 46    & 18    & 28 \\
\hline
    hadoop word count & 3     & 495   & 115   & 67    & 68 \\
\hline
    simple authentication & 2     & 306   & 47    & 14    & 21 \\
\hline\hline
Total & 6     & 997   & 208   & 99    & 117 \\
\hline
\end{tabular}\\
\end{SmallOut}\vspace*{-3ex}
\label{table:example}
\end{table}
\section{Evaluation}
\label{sec:evaluation}
We conduct evaluations to explore the research questions:

\begin{table*}[t]
\centering
\caption{The overall result and the result of only mappings.}\vspace*{-2ex}
\begin{SmallOut}
\begin {tabular} {|c|r|r|r|r|r|r|r|r|r||r|r|r|r|r|r|r|r|r|r|r|r|r|}
\hline
   \multirow{2}*[-2pt] {Name}  & \multirow{2}*[-2pt] {Total}  & \multirow{2}*[-2pt] {Clean}  &\multicolumn{7}{|c||}{\tool} &  \multicolumn{7}{|c|}{MA1} \\
   \cline{4-17}&&&\multicolumn{1}{|c|}{-} & \multicolumn{1}{|c|}{s}& \multicolumn{1}{|c|}{+} & \multicolumn{1}{|c|}{m}& \multicolumn{1}{|c|}{\%}& \multicolumn{1}{|c|}{error}& \multicolumn{1}{|c||}{solution}&\multicolumn{1}{|c|}{-} & \multicolumn{1}{|c|}{s}& \multicolumn{1}{|c|}{+} & \multicolumn{1}{|c|}{m}& \multicolumn{1}{|c|}{\%}& \multicolumn{1}{|c|}{error}& \multicolumn{1}{|c|}{solution}\\
\hline\hline
    \CodeIn{accumulo} & 28    & 4     & 23    & 1     & 0     & 11    & 45.8\% & 3.6   & 17.4  & 0     & 23    & 1     & 0     & 0.0\% & -1.3  & 0.2  \\
\hline
   \CodeIn{cassandra} & 161   & 89    & 71    & 1     & 0     & 43    & 59.7\% & 3.9   & 18.6  & 3     & 66    & 3     & 0     & 0.0\% & 0.0   & 0.2  \\
\hline
   \CodeIn{karaf} & 71    & 68    & 2     & 1     & 0     & 1     & 33.3\% & 96.0  & 26.0  & 0     & 3     & 0     & 0     & 0.0\% & 0.0   & 0.0  \\
\hline
    \CodeIn{lucene} & 91    & 78    & 9     & 4     & 0     & 2     & 15.4\% & 13.0  & 25.6  & 3     & 10    & 0     & 0     & 0.0\% & 6.3   & 0.6  \\
\hline
   \CodeIn{poi}   & 20    & 9     & 9     & 2     & 0     & 4     & 36.4\% & 16.4  & 41.7  & 1     & 9     & 1     & 0     & 0.0\% & 2.0   & 0.6  \\
\hline
\end{tabular}\\
\end{SmallOut}\vspace*{-1ex}
\label{table:overall}
\end{table*}
\begin{enumerate}[(RQ1)]
\item  How effectively does \tool resolve compilation errors that are introduced by updating libraries to newer versions
(Section~\ref{sec:evaluation:overall})?
\item What are the differences between the code updated by \tool and the code updated by programmers (Section~\ref{sec:evaluation:dif})?%
\item  How effectively does \tool migrate real code (Section~\ref{sec:evaluation:result:real})?
\item  How effective is \tool, if it uses only API mappings (Section~\ref{sec:evaluation:mapping})?
\end{enumerate}

For RQ1, our results show that \tool reduced compilation errors of 92.7\% tasks; eliminated the compilation errors of 32.4\% tasks; and do not increase the errors of tasks.

For RQ2, our results show that 39.3\% of our fully migrated tasks are identical to manual code changes.


For RQ3, our results show that \tool eliminated all the compilation errors of two real clients which were introduced by updating their libraries to a more recent version.

For RQ4, our results show that no task was migrated with only mappings.
%

More details of our evaluations are listed on our website:

\noindent
\begin{footnotesize}
\url{https://anonymous.4open.science/r/fa1c702b-5ccd-47c3-985e-ec9f2aefd949/}.
\end{footnotesize}

\subsection{RQ1. Overall Effectiveness}
\label{sec:evaluation:overall}
\subsubsection{Setup}
\label{sec:evaluation:overall:set}


Table~\ref{table:subject} shows the subjects of our evaluations. We select five widely used Java libraries: \CodeIn{accumulo} provides a sorted and scalable data storage; \CodeIn{cassandra} is a scalable database; \CodeIn{karaf} is an enterprise application runtime; \CodeIn{lucene} is a search engine; and \CodeIn{poi} is a library to manipulate Microsoft documents. All the projects are collected from the Apache Foundation, since they provide the archives of all their versions. Column ``FV'' lists the first versions. Column ``LV'' lists the last versions. Column ``TV'' lists the number of the total versions. We select all the versions of these libraries.

 All the five libraries provide API examples to illustrate how to call their APIs. Column ``LOC'' lists the total lines of API examples. Although API examples do not have many lines of code, they call many APIs of our subject libraries.
For example, Table~\ref{table:example} shows the examples of \CodeIn{cassandra 0.8.0-beta1}. Columns ``Class'', ``Method'', and ``Field'' show the numbers of unique called API classes, methods, and fields that are declared by \CodeIn{cassandra}. As shown in Table~\ref{table:example}, the three examples have only six files, but they call two hundred API classes of \CodeIn{cassandra}. The prior studies~\cite{linares2014mining,zhong2017empirical} show that most clients call a few hot spots APIs. As a comparison, most real projects do not call many APIs from a single library like \CodeIn{cassandra}. As for API migration, it is even more challenging to migrate examples than most real code.

Given the examples of the \CodeIn{v1} version and the next \CodeIn{v2} version, we build a project (\CodeIn{p}), in which the source files of \CodeIn{v1}'s examples are added to the source scope of \CodeIn{p} and the binary files of \CodeIn{v2} are added to the class path of \CodeIn{p}. The examples of \CodeIn{v1} call the APIs of \CodeIn{v1}, and the binary files of \CodeIn{v2} provide the APIs of \CodeIn{v2}. As a result, the \CodeIn{p} project illustrates a task, in which programmers migrate the examples from \CodeIn{v1} to \CodeIn{v2}. If \CodeIn{v2} contains breaking changes, the \CodeIn{p} project will have compilation errors. In such cases, programmers have to manually resolve all the compilation errors, if they migrate the examples from \CodeIn{v1} to \CodeIn{v2}. In this way, we construct a migration task for each version of a library.

We use \tool to migrate our tasks, and count the number of reduced compilation errors as the measure for the effectiveness of a migration tool. We select this measure, because reducing such errors is a critical step of migrating client code. Indeed, when programmers manually migrate projects to call newer libraries, they also have to resolve all compilation errors before they can test and debug bugs that are introduced in their migration process. We notice that some researchers (\emph{e.g.}, \cite{zhong2010mining}) also use this measure to analyze the effectiveness of their approaches. Although superficial modifications can resolve compilation errors, they can lead to semantic bugs in migrated code. As a result, fewer compilation errors do not necessarily indicate better migration results. To resolve this issue, we compare our migrated tasks with the existing migrate projects in Section~\ref{sec:evaluation:dif}. Furthermore, in Section~\ref{sec:evaluation:result:real}, we use \tool to migrate two real projects, and execute their test cases to ensure that its migrations are correct.

We did not compare \tool with prior approaches, because it is difficult to obtain change examples. For example, Xing and Stroulia~\cite{xing2007api} requires UML changes, but most libraries do not present their UMLs. As it is infeasible to align their inputs, any comparison is illustrative but not conclusive.



\subsubsection{Result}
\label{sec:evaluation:overall:result}


Table~\ref{table:overall} shows the result. Column ``Total'' shows the number of total tasks. Column ``Clean'' shows the number of tasks whose \CodeIn{p} projects have no compilation errors. These \CodeIn{p} projects have no compilation errors, since their API changes are all compatible. For the other \CodeIn{p} projects, Column ``\tool'' lists our results. Subcolumns ``-'', ``s'', and ``+'' show the number of the tasks where compilation errors of \CodeIn{p} projects are reduced, unchanged, and increased, respectively. \tool reduces the compilation errors for more than 90\% tasks, without increasing such errors of any tasks. Subcolumns ``migrated'' shows the number of tasks whose compilation errors are eliminated. In total, \tool fully migrated 61 tasks. Subcolumns ``\%'' are calculated as $\frac{migration}{total-clean}$. In total, \tool eliminates the compilation errors of 33.2\% tasks. Subcolumns ``error'' show the averages of reduced compilation errors per task. Subcolumns ``solution'' show the averages of generated solutions per tasks, before the best one is found.

From Table~\ref{table:overall}, we identify two types of API evolutions. In the first type of evolutions, most next versions have trivial breaking changes. As a typical example, in \CodeIn{accumulo}, 24 out of its 28 \CodeIn{p} projects have compilation errors. \tool is able to eliminate their compilation errors, for about half of the \CodeIn{p} projects, and it needs to resolve 3.6 compilation errors per task. In the second type of evolutions, fewer near releases have breaking changes, but if they have such changes, their changes are nontrivial. As a typical example, in \CodeIn{karaf}, 68 out of 71 \CodeIn{p} projects have no compilation errors, but the other three projects have hundreds of compilation errors. On average, for \CodeIn{karaf}, \tool resolved 96 compilation errors per task, which are much more than those for \CodeIn{accumulo} and \CodeIn{cassandra}, but it still fails to fully resolve the three projects.

Table~\ref{table:overall} highlights the importance of our migration algorithm. On average, \tool has to try more than twenty solutions, before it finds the best ones. Without a migration algorithm, if a tool resolves one error for each task, it can still reduce compilation errors in \CodeIn{p} projects, but it is unlikely to generate the best solutions for most tasks.

In summary, our results show that \tool reduces the compilation errors of more than 90\% tasks, and 32.4\% tasks are fully resolved. \tool is more suitable to update client code whose libraries have continuous and trivial changes, but its effectiveness is reduced when handling libraries with sudden and nontrivial changes. In addition, most best solutions are not synthesized by single migration operators, but the combinations of them.

\subsection{RQ2. Comparison with Manual Migrations}
\label{sec:evaluation:dif}
\subsubsection{Setup}
\label{sec:evaluation:dif:set}

In total, 61 tasks were fully migrated. We manually inspected all these tasks to understand the differences between manual migrations and those of our tool. In each task, a \CodeIn{p} project is built by combining the example of \CodeIn{v1} and the binary files of \CodeIn{v2}. We checked whether the examples of \CodeIn{v2} and its later versions have the same compilation errors. If programmers migrated such errors, we checked whether their modifications are consistent with what \tool did on the \CodeIn{p} projects. Here, we ignore superficial edits (\emph{e.g.}, refactoring) and edits on comments.
\subsubsection{Result}
\label{sec:evaluation:dif:result}
Based on our inspection results, we classified the tasks into four categories:

\emph{1. In  24 tasks, our migrated projects are identical to manual migrations.} For example, the \CodeIn{p} project of \CodeIn{cassandra} \CodeIn{1.0.0} has an error: ``The method hexToBytes(String) is undefined for the type FBUtilities''. \tool replaces the method call with \CodeIn{hexToBytes(String)} of the \CodeIn{By\-te\-Buffer\-Util} type, which is identical to manual migrations.

\emph{2. In 33 tasks, we cannot compare our migrated projects with manual migrations, since their compilation errors are never resolved by programmers.} Typically, API tutorials do not provide any APIs. As programmers can pay less attention in updating examples, we find that these compilation errors are never resolved. For example, the \CodeIn{p} project of \CodeIn{cassandra} \CodeIn{0.8.0} includes the
\CodeIn{Cas\-san\-dra\-Bulk\-Loa\-der} type that calls the constructor:

\begin{lstlisting}
baseColumnFamily = new ColumnFamily(ColumnFamilyType.Standard, DatabaseDescriptor.getComparator(keyspace, columnFamily), DatabaseDescriptor.getSubComparator(keyspace, columnFamily), CFMetaData.getId(keyspace, columnFamily));
\end{lstlisting}

As the constructor is deleted in the later versions, the above code produces a compilation error. The most recent version of the type appears in \CodeIn{cassandra} \CodeIn{0.8.10}, but even in that version, this compilation error is not removed. \tool replaces the call with a creator to resolve the problem:

\begin{lstlisting}
baseColumnFamily = ColumnFamily.create(CFMetaData.getId(keyspace, columnFamily));
\end{lstlisting}

\emph{3. In two tasks, the buggy code is deleted.} For example, the \CodeIn{p} project of \CodeIn{cassandra} \CodeIn{1.2.0-beta3} has an example, in which a code line is as follows:

\begin{lstlisting}
EnumSet<Permission> authorized = Permission.NONE;
\end{lstlisting}

As the type of \CodeIn{Permission.NONE} is modified to \CodeIn{Set\-$<$Permis\-sion$>$} in the later versions, the above code example produces a type mismatch. \tool adds an explicit cast to resolve it:

\begin{lstlisting}
EnumSet<Permission> authorized = (EnumSet<Permission>)Permission.NONE;
\end{lstlisting}

We found that the later versions delete this example, so we cannot compare our modifications with manual migrations.

\emph{4. In two tasks, API code is modified to resolve their compilation errors.} For example, the \CodeIn{p} project of \CodeIn{lucene} \CodeIn{4.0.0} has an error: ``The constructor TextField(String, BufferedReader) is undefined'', since its next version deletes this constructor. \tool modified client code to resolve this issue, but \CodeIn{lucene} programmers roll back the deletion to resolve the issue, and their client code is not changed.

In summary, we find that 39.3\% migrated tasks are identical to manual modifications, and in 54.1\% tasks, compilation errors are ignored by programmers but are resolved by \tool. In four tasks, programmers present other ways to resolve update issues, \emph{e.g.}, updating API code.
\begin{figure}[t]
\centering
\includegraphics[scale=0.5,clip]{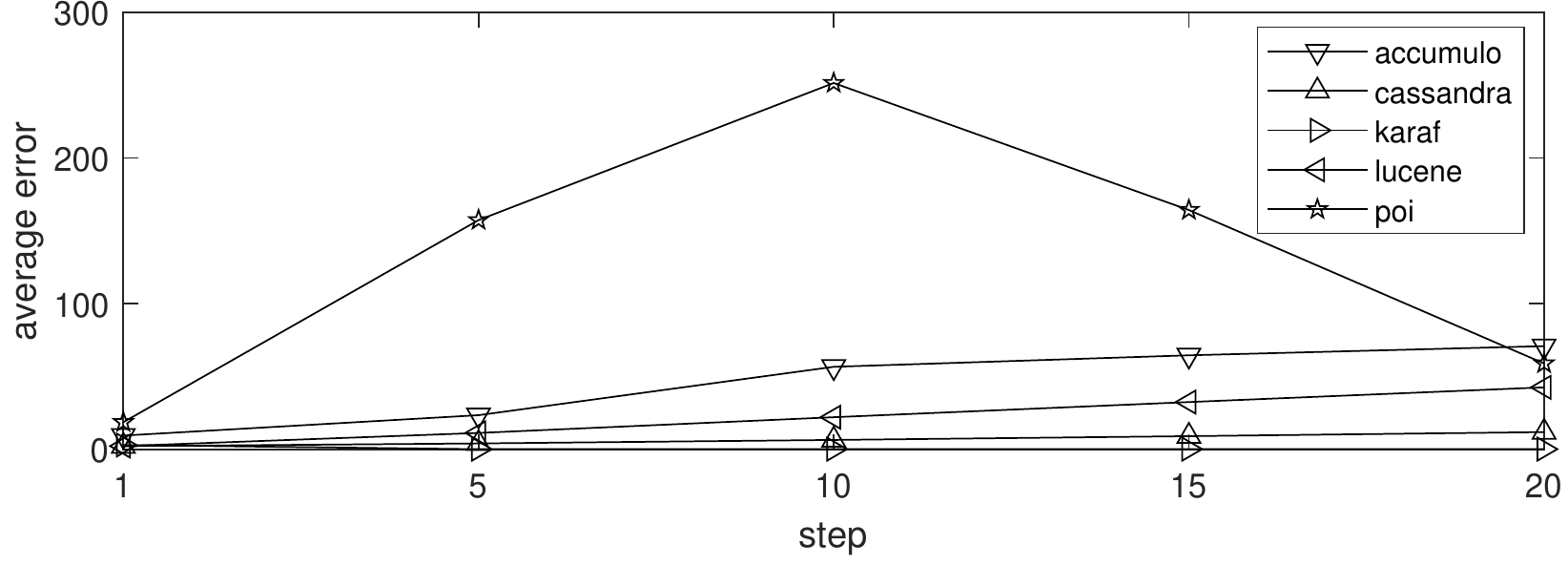}\vspace*{-2ex}
 \caption
{\label{fig:averageerror} The average errors}\vspace*{-2ex}
\end{figure}

\subsection{RQ3: Real Migrations}
\label{sec:evaluation:result:real}
\subsubsection{Setup} In this section, we focus on the clients of \CodeIn{lucene}, since it is a popular library. In total, there are 427 projects under the \CodeIn{lucene} topic of Github~\cite{luceneclient}.
Among those projects, we included projects into our evaluation if
any project $P$ satisfies all of the following five criteria:
(1) $P$ uses any of \CodeIn{lucene}'s versions lower than \CodeIn{7.4.0}; (2) $P$ uses Maven as the build system, because Maven allows us to easily change the version information of any library dependency; (3) $P$ can be successfully built without any compilation error; (4) $P$ is not a toy project according to the project's description, and we also require that $P$ should be  forked and have stars; and (5) $P$ has unit test cases because after migration, we run those test cases to detect possible errors. With this filtering process, we ended up with two client projects: \CodeIn{ESA} and \CodeIn{FleaDB}.

According to semantic versioning~\cite{sv}, the first number of a version indicates that this version makes incompatible API changes. For each selected project, we changed the version of its \CodeIn{lucene} library (\CodeIn{v1}) to the next version whose first number is different from \CodeIn{v1}. This setting analyzes to what degree \tool can resolve such incompatible API changes.
\subsubsection{Updating \CodeIn{ESA}}
\CodeIn{ESA}~\cite{esa} is the abbreviation of Explicit Semantic Analysis. It leverages Wikipedia dumps to compare the semantic similarity between two given texts. It is a research tool, and its papers are published in the top AI venues~\cite{gabrilovich2007computing,gabrilovich2006overcoming}. Google scholar reports that the two papers have more than 3,000 citations. The latest \CodeIn{ESA} has 1,514 lines of code, and is built on \CodeIn{lucene} \CodeIn{4.8.1}.

In this migration task, we replaced its \CodeIn{lucene} library from \CodeIn{4.8.1} to \CodeIn{5.0.0}. After that, \CodeIn{ESA} produced 23 compilation errors. These compilation errors include 9 unresolved variables, 6 undefined constructors or methods, 6 incompatible parameters, and 2 unimplemented methods. For example, an error message is ``LUCENE\_48 cannot be resolved to a variable'', because  \CodeIn{LUCENE\_48} is deleted in \CodeIn{5.0.0}.

\begin{figure}[t]
\centering
\includegraphics[scale=0.5,clip]{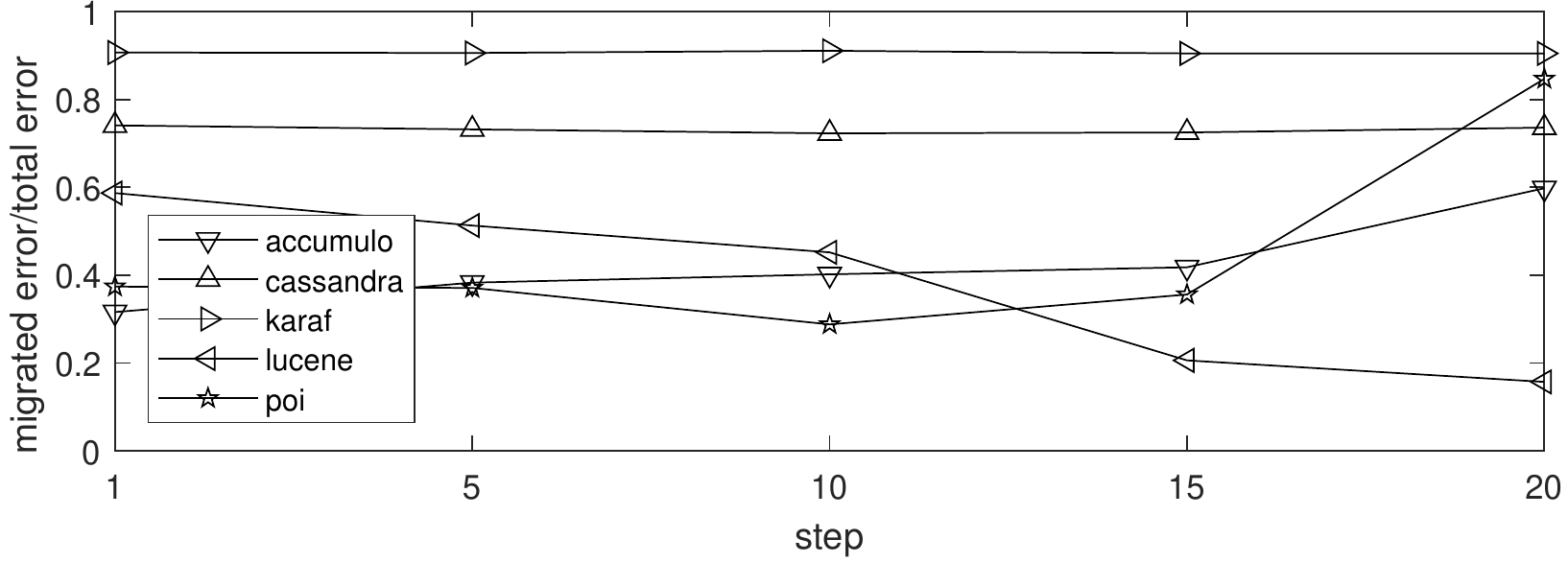}\vspace*{-1ex}
 \caption
{\label{fig:errorratio} The ratios of migrated errors}\vspace*{-2ex}
\end{figure}

\tool migrated \CodeIn{ESA} in only several minutes, and all the compilation errors are resolved. After inspecting the migration log, we find that MA6 resolved 16 compilation errors. For example, MA6 made a modification as follows:

\begin{lstlisting}
-queryParser=new QueryParser(LUCENE_48, TEXT_FIELD, analyzer);
+queryParser=new QueryParser(TEXT_FIELD, analyzer);
\end{lstlisting}

When \CodeIn{lucene 5.0.0} removes the \CodeIn{LUCENE\_48} constant, it also removes the first formal parameter of the \CodeIn{QueryParser} constructor. MA6 removed the actual parameter of the old version to resolve this compilation error.

MA8 resolved 5 compilation errors, and one is as follows:
\begin{lstlisting}
-... = FSDirectory.open(termDocIndexDirectory)){
+... = FSDirectory.open(termDocIndexDirectory.toPath()){
\end{lstlisting}

In the above patch, the type of \CodeIn{termDocIndexDirectory} was \CodeIn{File}, but the type of the formal parameter was changed to \CodeIn{Path}. MA8 resolved the problem by calling the \CodeIn{toPath()} method of the actual parameter. Besides the above modifications, MA9 generated stubs to resolve the remaining 2 unimplemented methods. \CodeIn{ESA} has two test methods, and our migrated code passed both test methods.

\subsubsection{Updating \CodeIn{FleaDB}} \CodeIn{FleaDB}~\cite{fleadb} is a database that supports the persistence of objects. It has 2,611 lines of code, and is built on \CodeIn{lucene} \CodeIn{4.10.3}. The programmer of \CodeIn{FleaDB} explains that the APIs of \CodeIn{lucene} often change and \CodeIn{FleaDB} can provide more stable APIs.

In this migration task, we also replaced its \CodeIn{lucene} library with \CodeIn{5.0.0}. The replacement introduced 6 compilation errors, including 2 undefined methods and 4 incompatible parameters. \tool resolved all the six compilation errors. After inspecting its migration log, we find that MA6 resolved 3 compilation errors; MA8 resolved 2 compilation errors, and MA11 resolved 1 compilation error. For example, a modification is as follows:

\begin{lstlisting}
-TopFieldCollector.create(this.sort, 1, null, false, false, false, false);
+TopFieldCollector.create(this.sort, 1, false, false, false, false);
\end{lstlisting}

In \CodeIn{5.0.0}, the third formal parameter of the \CodeIn{create} method is deleted. MA8 removed the corresponding actual parameter to solve the problem. \CodeIn{FleaDB} implements 15 test methods, and our migrated code passed all the test methods.

In summary, \tool migrated two projects to call a more recent version of \CodeIn{lucene}, and its migrated code passed all the test methods of both projects. The results show the significance of \tool, in that some migration edits are complicated as shown in the above examples.

\subsection{RQ4. Migrating with Only API Mappings}
\label{sec:evaluation:mapping}
\subsubsection{Setup}
\label{sec:evaluation:mapping:set}
As Lamothe and Shang~\cite{lamothe2018EUA} report that many approaches mine API mappings, it is interesting to explore the effectiveness of our approach, if migrates code with only such mappings. In this research question, we enable only MA1 of \tool to update code.

\subsubsection{Result}
\label{sec:evaluation:step:result}

In Table~\ref{table:overall}, Column ``MA1'' lists the migration results of only mappings. Compared with Column ``\tool'', MA1 does not change the compilation errors for more than 90\% tasks, and for five tasks, it even increases such errors. Although MA1 reduces the compilation errors of seven tasks, it fails to fully migrate any task. As a comparison, \tool resolves more compilation errors and generates more solutions than MA1 does per task.

\subsection{Threats to Validity}
\label{sec:evaluation:threat}
The threats to external validity include our subjects. We select only Apache projects, since their websites provide all their releases. To improve the generality of our results, a tool needs to select projects from more sources, if such projects also provide the archives of all the releases and provide code examples to illustrate their API usages. The construct threat to validity includes that as programmers can forget to update their API examples, the compilation errors in our migration tasks are more than the cases where client code is up-to-date. As a result, the true effectiveness of \tool can be underestimated. The problem can be mitigated by migrating client code from more reliable sources. The threat to interval validity includes the parameters of our migration algorithm, which can be less optimized and can be improved.

\section{The Potential of Our Direction}
\label{sec:discuss}

Our work opens a new research direction to migrate obsolete client code, and our evaluation results show that for the first time, our tool is able to migrate real projects without change examples. As it is tedious and even infeasible to find change examples for all possible migrations, our achievement is a critical step towards practical migration tools. Still, there is still sufficient space for improvements, and the space can attract and incubate a small research community. Besides what were already discussed in Section~\ref{sec:road}, some other research opportunities are as follows:

\emph{1. Migrating other languages.} Another programming language can call APIs in a way that is different from Java or other object-oriented languages, and its migration process can involve more issues. To enable the migration on other languages, researchers shall design other migration operators, and the migration algorithm can be modified according to its new features.

\emph{2. Tuning our treatments and independent variables.} Tuning the treatments and independent variables of \tool can squeeze its last bit out. For example, trying other search algorithms~\cite{price1983global,schutte2004parallel} and tuning our parameters~\cite{angelova2011tuning} can improve our effectiveness. As another example, when a driver produces a compilation error, Lawall \emph{et al.}~\cite{lawall2017fast} propose an approach to locate its related patches of kernels, and similar techniques can recommend better migration operators.


\section{Related Work}
\label{sec:related}

\textbf{Code migration.} With evolution of libraries, some APIs
may become incompatible across library versions. To address this
problem, Henkel and Diwan~\cite{henkel2005catchup} proposed an approach that captures
and replays API refactoring actions to update the client code.
Xing and Stroulia~\cite{xing2007api} proposed an approach that
recognizes the changes of APIs by comparing the differences between two
UMLs of libraries. Kalra \emph{et al.}~\cite{Kalra2016PSU} match traces of client code in two versions for their mappings. Balaban \emph{et al.}~\cite{balaban2005refactoring} proposed
an approach to migrate client code when mapping relations of libraries are already
available. Xu \emph{et al.}~\cite{xu2019meditor} learn edit scripts from migration commits and apply them on new migration tasks. \tool is the first approach to combine simple edits to complicated migrations.

\textbf{Search-based software engineering.} Harman \emph{et al.}~\cite{harman2012search} conducted a review on the search-based software engineering. As a typical search algorithm, genetic algorithms have been widely used in various applications such as generating test cases~\cite{srivastava2009application}, release planning~\cite{greer2004software}, repairing bugs~\cite{weimer2009automatically}, refactoring~\cite{o2008search}, and estimating development cost~\cite{huang2006optimization}. Our migration algorithm is inspired by search algorithms, but it resolves in a different research problem, complementing the above approaches.

\textbf{Learning transformations from change examples.} Given an original file and its modified file, Andersen \emph{et al.}~\cite{andersen2010generic} extract a set of term replacements from change examples. Meng \emph{et al.}~\cite{MengKM11,Meng:LASE} learn edit scripts from change examples. Rolim \emph{et al.}~\cite{rolim2017learning} search for a transformation that is consistent with all change samples. Long \emph{et al.}~\cite{long2017automatic} infer AST templates from patches. Nguyen
\emph{et al.}~\cite{nguyen2019graph} mine graph change patterns from change examples. Given a change example, Jiang \emph{et al.}~\cite{jiang2019inferring} mine where to apply its transformation. Given change examples illustrate migrations, Fazzini \emph{et al.}~\cite{fazzni2019issta} learn transformations to migrate code. Zhang \emph{et al.}~\cite{zhang2019large} conduct an empirical study on compiler errors in continuous integration. Mesbah \emph{et al.}~\cite{mesbah2019deepdelta} learn a model from past fixes to repair compilation errors. \tool is not a learn-based approach, since it does not require change examples as its inputs.

\section{Conclusion}
\label{sec:conclusion}

The prior approaches typically migrate client code with migration samples, but it is difficult to obtain such samples. In this paper, we propose \tool. Compared with the state of the art, \tool presents a new research direction, in which client code can be migrated without migration samples. Its basic idea is to combine simple edits of migration operators under its migration algorithm, until a complex migration is synthesized. Our evaluation results show that \tool resolves all the compilation errors of 32.4\% migration tasks; 39.3\% migrated tasks are identical to manual migrations; and two real projects were fully migrated.

\balance
\bibliographystyle{abbrv}
\bibliography{autoupdate}
\end{document}